\documentclass[twocolumn,english,superscriptaddress,pre]{revtex4-2}
\usepackage[T1]{fontenc}
\usepackage[latin9]{inputenc}
\usepackage{amsmath}
\usepackage{amssymb}
\usepackage{babel}
\usepackage[colorlinks=true,allcolors=blue]{hyperref}
\usepackage{xspace}
\usepackage{xcolor}
\usepackage{siunitx}
\usepackage{physics}
\usepackage{graphicx}
\graphicspath{{figures/}} 

\newcommand{\Ns}{N_{\text{s}}}
\newcommand{\Nc}{N_{\text{c}}}
\newcommand{\Np}{N_{\text{p}}}
\newcommand{\NA}{N_{\text{A}}}
\newcommand{\Vi}{V_{\text{i}}}
\newcommand{\Veq}{V_{\text{eq}}}
\newcommand{\phii}{\phi_{\text{i}}}
\newcommand{\phieq}{\phi_{\text{eq}}}
\newcommand{\nus}{\nu_{\text{s}}}
\newcommand{\nup}{\nu_{\text{p}}}
\newcommand{\kb}{k_{\text{B}}}
\newcommand{\boldf}[1]{\textbf{#1}}
\begin{document}
\title{Morpho---A programmable environment for shape optimization and shapeshifting problems}
\author{Chaitanya Joshi}
\affiliation{Tufts University Department of Physics and Astronomy, 574 Boston Ave,
Medford, MA 02155}
\author{Daniel Goldstein}
\affiliation{Tufts University Department of Physics and Astronomy, 574 Boston Ave,
Medford, MA 02155}
\author{Cole Wennerholm}
\affiliation{Tufts University Department of Physics and Astronomy, 574 Boston Ave,
Medford, MA 02155}
\author{Eoghan Downey}
\affiliation{Tufts University Department of Physics and Astronomy, 574 Boston Ave,
Medford, MA 02155}
\author{Emmett Hamilton}
\affiliation{Tufts University Department of Physics and Astronomy, 574 Boston Ave,
Medford, MA 02155}
\affiliation{Tufts University Department of Mathematics, 177 College Ave, Medford,
MA 02155}
\author{Samuel Hocking}
\affiliation{Tufts University Department of Mathematics, 177 College Ave, Medford,
MA 02155}
\author{Anca Andrei}
\affiliation{Tufts University Department of Mathematics, 177 College Ave, Medford,
MA 02155}
\author{James H. Adler}
\affiliation{Tufts University Department of Mathematics, 177 College Ave, Medford,
MA 02155}
\author{Timothy J. Atherton}
\affiliation{Tufts University Department of Physics and Astronomy, 574 Boston Ave,
Medford, MA 02155}

\date{\today}%

\begin{abstract}
    An emerging theme across many domains of science and engineering is materials that change shape, often dramatically. Determining their structure involves solving a shape optimization problem where a given energy functional is minimized with respect to the shape of the domain and auxiliary fields describing the structure. Such problems are very challenging to solve and there is a lack of suitable simulation tools that are both readily accessible and general purpose. To address this gap, we present \emph{Morpho}, an open-source programmable environment, and demonstrate its versatility by showcasing three applications to different areas of soft matter physics---swelling hydrogels, complex fluids that form aspherical droplets, to soap films and membranes---and advise on broader uses. 
\end{abstract}

\maketitle


Numerous domains of scientific research involve the solution of shape optimization and evolution problems: soft robots able to change their shape \cite{Wehner2016,Booth2018,Cunha2020,Shah2021,Mengaldo2022}; complex fluids, particulate matter and liquid crystals with free boundaries \cite{Kaplan2010,Xing2012a,Giomi2012b,Gibaud2017,Ding2021,Carenza2022,Khanra2022,Safdari2021}; systems where mechanical properties emerge from the structure \cite{Silverberg2014,Silverberg2015,Paulose2015}; active biological materials and membranes that are internally driven \cite{Keber2014,Giomi2014, Maroudas-Sacks2020,Vutukuri2020,Peterson2021,Khoromskaia2021,Hoffmann2022}; multiphase systems; gels \cite{Jensen2015,Datta2016}; responsive polymeric materials \cite{SydneyGladman2016a,Na2015,Silverberg2014}, computational differential geometry \cite{goodman-strauss_sullivan_2003}, etc. to name a few. Cutting across many of these applications is the idea of shape as a programmable quantity \cite{Hawkes2010,Na2015,Silverberg2014} where a material or system is guided to a desired final configuration by local structure and an applied external influence such as a magnetic field or a chemical cue. In each of these scenarios, some energy functional must be minimized with respect to the shape of the system of interest together with other quantities such as electric or magnetic fields and also subject to imposed constraints. 

In this Article, we introduce a programmable environment, \emph{Morpho}, that aims to solve the following class of problems. Suppose we have a domain, $C=\bigcup_{i}c_{i}$ composed of subdomains $c_{i}$, in $n$ dimensional space. On $C$ are defined zero or more scalar or tensor fields, $\mathbf{q}$. The goal is to optimize a given energy functional with respect to the shape of the domain and the field values, 
\[
F=\sum_{i}\int_{c_{i}}f_i(\mathbf{q},\nabla \mathbf{q}, ...)d^{n}x+\sum_{i}\int_{\partial c_{i}}g_i(\mathbf{q}, \nabla \mathbf{q},...)d^{n-1}x,
\]
subject to global (integral) constraints,
\[
\sum_{i}\int_{c_{i}}h_{i}(\mathbf{q},\nabla \mathbf{q})d^{n}x=0,
\]
together with local constraints $u_{k}(\mathbf{q},\nabla \mathbf{q})=0$. Here, $f_i$, $g_i$, $h_i$ and $u_k$ are scalar functions of $\mathbf{q}$ and its gradients; these may differ by subdomain. The constraint set could also include inequality constraints such as limitations on the shape including mutual inter-penetrability between elements or excluding the mesh from encroaching on a certain region. 

Numerous important problems lie in the class just described. For example, the shape of a soap film locally minimizes the area due to surface tension. In a compact geometry, i.e., a soap bubble, the volume may either be considered as fixed (implying a global constraint) or established by balancing the pressure difference inside and outside the droplet against the surface force. Non-compact geometries also can occur, where the shape of the boundary is prescribed, as in a soap film suspended by a wire. 

We briefly review a number of other techniques that are used to solve shape optimization problems. Phase field methods \cite{provatas_elder_2010,Chen2002} use an auxiliary scalar field called the phase-field, that smoothly interpolates between the interior and exterior of a shape. Phase field methods are straightforward to implement, but resolving sharp features such as cusps that may arise in such problems can be difficult. Level set methods \cite{sethian_2010} represent the free boundary of the system as a contour or a level set of a scalar function defined in a higher dimensional space. An advantage of this is that changes in topology, such as the coalescence of two fluid droplets, can be easily handled. Formulating the optimization problem for a particular problem requires sophisticated techniques, however, and enforcement of constraints can be challenging. 

In contrast, \emph{Morpho} uses an explicit discretization of the  problem domain and associated quantities. From these, \emph{Morpho} is able to evaluate the objective function of interest as well as its gradients and Hessian with respect to mesh and field degrees of freedom. A number of algorithms for constrained optimization are then available within the \emph{Morpho} environment to perform the optimization using these quantities. The approach is similar in spirit to the highly successful \emph{Surface Evolver} (SE) software \cite{Brakke1992,Brakke1996}. Originally designed for minimal surface computations, SE has been adapted to many other uses. Surface Evolver's success in part comes from its interactive, re-programmable and versatile nature, with direct applicability to arbitrary problems. Nonetheless, SE works with surfaces only and does not support optimization with respect to field quantities. Further, mesh quality control in SE requires user intervention, including manual refinement and coarsening steps. We significantly address these limitations in \emph{Morpho}. 

\begin{figure*}
    \centering
    \includegraphics[width=0.9\textwidth]{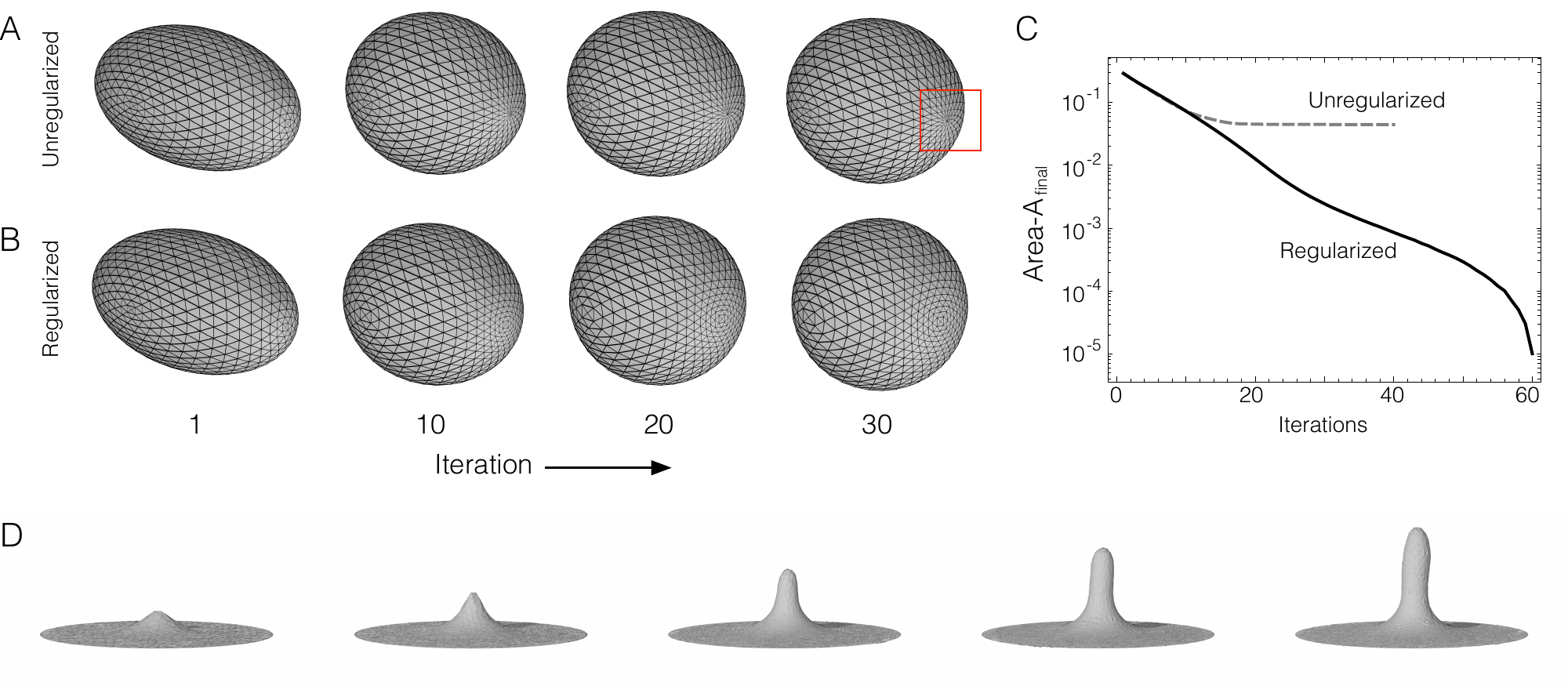}
    \caption{\boldf{Soap films and membranes}. Optimization of a soap film from an initial ellipsoidal configuration under surface tension at fixed enclosed volume: \boldf{A} without regularization and \boldf{B} with regularization. In the absence of regularization, vertices near the end of the solution become bunched together (highlighted region). \boldf{C} Residual area as a function of iteration showing that the unregularized solution converges on an incorrect solution. \boldf{D} Simulated drawing of a tether from a circular disk of bilayer membrane under the Helfrich energy. }
    \label{fig:membranes}
\end{figure*}

From the user's perspective, \emph{Morpho} provides a very flexible, object-oriented environment for shape optimization. A user may work with the program interactively or by supplying a script to be run, written in a simple but powerful dynamic programming language similar to \emph{Python}. The language is modular, thus supporting user-defined libraries and customization, and enables specification of a problem to closely correspond to the mathematical problem statement. Documentation is available through online help in interactive mode, a manual and a website. Convenient interactive visualization is provided through a companion application \texttt{morphoview}.

In the remainder of the paper, we describe a number of illustrative example applications of \emph{Morpho}. First, we solve a classic problem, the shape of a soap film under surface tension at fixed volume. This example illustrates a key innovation of our work, which is the use of \emph{auxiliary} functionals to automatically control mesh quality, improving the quality and accuracy of the solution. We then use \emph{Morpho} to resolve the shape of anisotropic liquid crystal fluid droplets or \emph{tactoids} in 2D and 3D, illustrating combined shape and field minimization. Finally, we study hydrogels swelling under confinement to illustrate the sophisticated use of constraints. We conclude with a discussion of \emph{Morpho}'s feature set, its potential applicability to other research areas and avenues for future development. Further details of the environment and its implementation, together with additional information about the applications presented here are then provided in Methods. 

\section{Results}

\subsection{Example 1: Minimal surfaces and membranes}

An important class of problems include finding surfaces that minimize an energy functional. As we note in the introduction, a soap film minimizes the area,
\begin{equation}
F=\int_{C}dA,\label{eq:Area}
\end{equation}
where the domain $C$ is a surface embedded in $\mathbb{R}^{3}$. Because the minimizer of (\ref{eq:Area}) vanishes to a point, typical problems involving soap films involve constraints. For example, the total volume enclosed in a compact soap film may be fixed, or the boundary may be specified. A related problem is to find the shape of a membrane, such as that enclosing a biological cell, which minimizes the Helfrich energy \cite{Canham1970,Helfrich1973,Kamien2002}, 
\begin{equation}
F=\sigma\int_{C}dA+\kappa\int_{C}\left(H-H_{0}\right)^{2}dA+K_{G}\int_{C}KdA+\lambda\int_{\partial C}dl,
\label{eq:Helfrich}
\end{equation}
where $H$ is the local mean curvature, $H_{0}$ is a locally preferred mean curvature, $K$ is the Gaussian curvature and $\sigma$, $\kappa$, $K_{G}$ and $\lambda$ are (material dependent) constants. 

While these problems are quite simple to state, they enable us to explore some of the numerical subtleties of shape optimization problems before turning to more complex applications. An ubiquitous challenge is that the initial geometry specified by the user may prove to be very different from the solution, which could have a different topology, for example, or include features that are not hinted at in the starting point. Hence, shape optimization necessitates some form of adaptive mesh control. Strategies to maintain mesh quality include splitting elements where the solution is poorly resolved, merging elements where there is excessive detail, or redistributing elements to improve their density in regions of interest. \emph{Morpho} provides support for all of these. 

As an illustration, we show in Fig. \ref{fig:membranes} an initially ellipsoidal mesh that is relaxed back to a sphere under surface tension using a gradient descent scheme with line searches (See Methods). Figs.  \ref{fig:membranes}A and  \ref{fig:membranes}B show intermediate snapshots of the system in the first few iterations with and without mesh control, respectively; Fig. \ref{fig:membranes}C shows the area of the system as a function of iteration number. As is evident, without mesh control, the system converges spuriously on an incorrect solution. Inspection of the solution (see red highlighted region in Fig.  \ref{fig:membranes}A) reveals why this occurs: mesh points at the ends of the ellipsoids are becoming bunched together. To explain this, we use an elementary result from differential geometry, that the gradient of the area with respect to a point on the surface locally lies in the direction of the surface normal. Hence, the target problem is under-determined; the tangential distribution of the vertices is not fully specified by the functional of interest.

To repair the situation, we supplement the problem (\ref{eq:Area}) with an auxiliary \emph{regularization} problem that penalizes differences in area between adjacent elements. This new problem is co-minimized together with the original problem; in practice we find only occasional regularization steps are necessary. With such steps, the algorithm converges correctly on a spherical solution as seen in Fig. \ref{fig:membranes}C. In more complex problems, the continuous regularization scheme presented here can be supplemented with occasional discrete refinement steps, i.e., splitting or merging elements. 

As an illustration of the use of \emph{Morpho}'s mesh control features to solve a more challenging problem, in Fig. \ref{fig:membranes}D we show the shape and formation of a tether from an initial disk-shaped patch of lipid membrane with fixed boundary. Such tethers occur in many biologically-relevant scenarios involving lipid bi-layers including micro-manipulation experiments on artificial vesicles as well as the Golgi apparatus\cite{powers2002fluid}. The disk is subject to localized indentation, drawing out a cylindrical tether beyond a certain displacement. 

\subsection{Example 2: Liquid crystal tactoids}

\begin{figure*}
    \centering
    \includegraphics[width=0.9\textwidth]{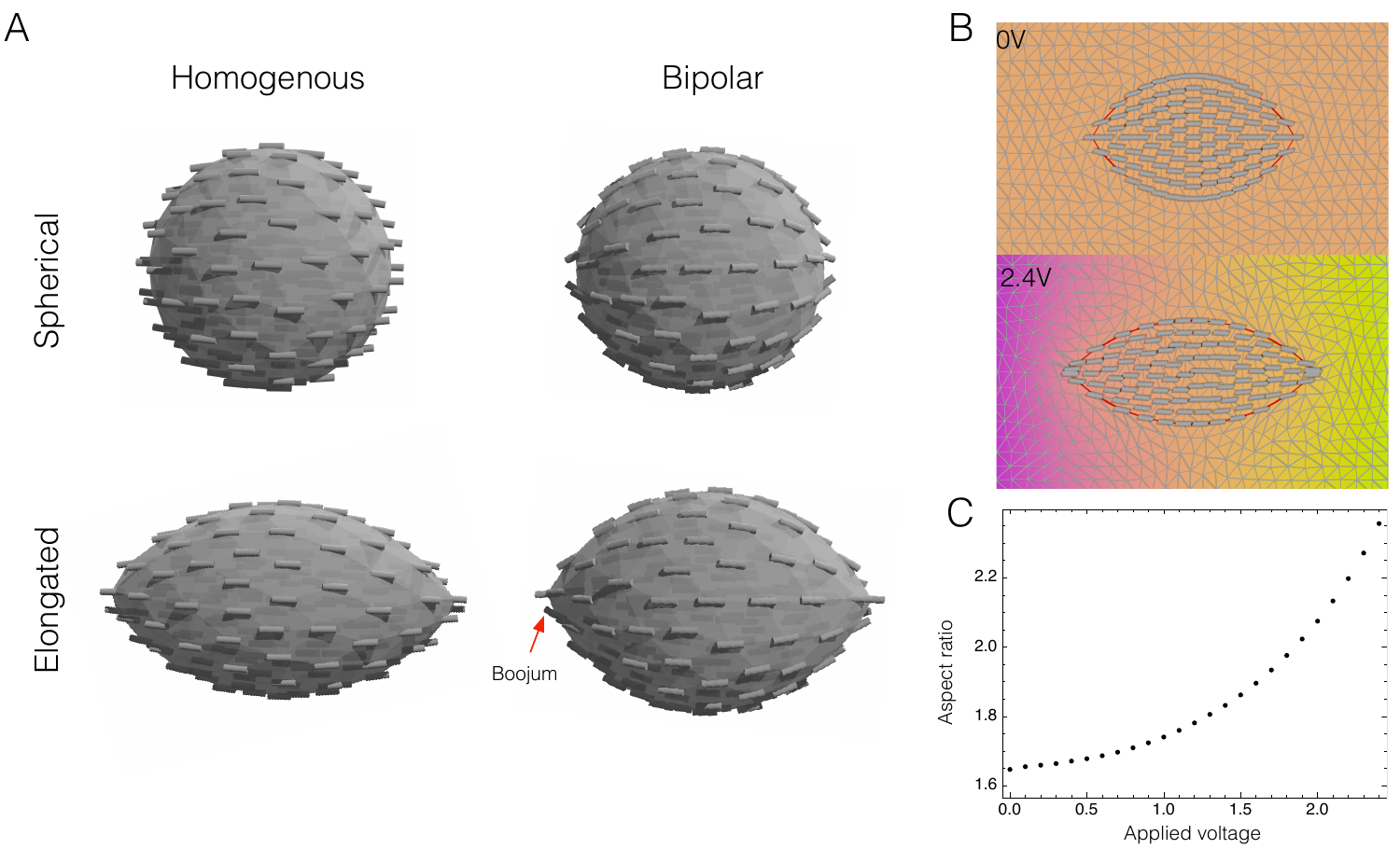}
    \caption{\boldf{Nematic liquid crystal tactoids.} \textbf{A} Three dimensional tactoid morphologies that occur in various parameter regimes. \textbf{B} Initial and final 2D tactoid shapes in the presence of an electric field. \textbf{C} Equilibrium aspect ratio as a function of applied voltage. }
    \label{fig:tactoid}
\end{figure*}

In contrast to isotropic fluids, anisotropic fluids can support elastic deformation, orientationally dependent surface tension and other physical effects that make non-spherical equilibrium droplet shapes possible. A commonly-encountered example of such a fluid is a nematic liquid crystal (NLC), which is composed of long, rigid molecules that tend to align locally in some preferred direction. These materials are commercially important for display and electro-optic applications, as well as emerging technologies such as chemical and biological sensors.

One way of mathematically describing these materials is through a unit vector field $\mathbf{n}(\vb{r})$ known as the director field, which specifies the average direction in which the molecules align at a location $\vb{r}$. To determine the equilibrium shape of a droplet $C$, the functional to be minimized in three dimensions is the free energy,

\begin{align}
F= & \int_{C}\left[ \frac{1}{2}K_{11}\left(\nabla\cdot\mathbf{n}\right)^{2}+K_{22}(\mathbf{n}\cdot\nabla\times\mathbf{n})^{2} \right. \nonumber  \\
&\quad +\left. K_{33}\left|\mathbf{n}\times\nabla\times\mathbf{n}\right|^{2} \right] \dd V \nonumber \\
 & +\sigma\int_{\partial C} \dd A-\frac{W}{2}\int_{\partial C}\left(\mathbf{n}\cdot\mathbf{t}\right)^{2} \dd A,\label{eq:frank}
\end{align}

where the various contributions to the energy are as follows: the first term corresponds to liquid crystal elasticity, with three constants $K_{11}$, $K_{22}$ and $K_{33}$ measuring the cost of splay, twist and bend deformations, respectively \cite{degennes_prost_1993}; the second term is the surface tension with associated constant $\sigma$; and the final term imposes a preferred orientation at the boundary, a phenomenon known as anchoring. If the anchoring coefficient, $W>0$, the director, $\mathbf{n}$, prefers to align with the local tangent to the surface, $\mathbf{t}$.

The functional (\ref{eq:frank}) is to be minimized with respect to the shape of the domain $C$ and the configuration of the director field $\mathbf{n}$ subject to a volume constraint, 
\[
\int_{C} \dd V=V_{0},
\]
and a local constraint, 
\[
\mathbf{n}\cdot\mathbf{n}=1.
\]

By introducing a length scale derived from the volume of the droplet $V^{1/3}$ and defining a mean elastic constant $\bar{K}=\frac{1}{3}\sum K_{ii}$, the above expression can be nondimensionalized. The solution is then a function of some dimensionless parameters $\kappa=\bar{K}/\sigma V^{1/3}$, the ratio of elastic forces to surface tension, $\omega=W/\sigma$, the ratio of anchoring energy to surface tension, and the reduced elastic constants $k_{ii}=K_{ii}/\bar{K},i\in\{1,2,3\}$. 

In Figure \ref{fig:tactoid}A, we show paradigmatic solutions that illustrate the variety of possible morphologies. The local orientation of the director field is depicted using cylinders. For $\kappa\ll1$, the surface tension holds the droplet in a spherical shape; as $\kappa\gtrapprox1$ the droplet elongates to form a spindle shape. The director field also undergoes a transition: for $\omega<\omega_{o}(\kappa)$ elasticity overcomes anchoring leading to a homogenous director field; for $\omega>\omega_{0}(\kappa)$ the director aligns with the surface producing a bipolar configuration. The critical anchoring energy for these transitions has been predicted from a scaling theory \cite{Prinsen2003}; here we obtain the solutions by direct optimization. 

As well as elastic anisotropy, liquid crystals also exhibit \emph{dielectric anisotropy}, whereby the dielectric tensor $\epsilon=\epsilon_{\perp}I_{3}+\epsilon_{a}\mathbf{n}\otimes\mathbf{n}$ depends on the local orientation of the director field $\mathbf{n}$. Here, $\epsilon_{\perp}$ is the component of the dielectric tensor perpendicular to $\mathbf{n}$, $I_{3}$ is the $3\times3$ identity matrix and $\epsilon_{a}=\epsilon_{\parallel}-\epsilon_{\perp}$ is the dielectric anisotropy. The dielectric anisotropy has a number of physical consequences. At low frequency, if $\epsilon_{a}>0$ the director tends to reorient to align with an applied electric field. At optical frequencies, dielectric anisotropy implies birefringence. The combination of electrical switchability and optical activity facilitates the creation of switchable electro-optic devices such as displays. 

To predict the effect of an electric field on the shape of a droplet of nematic, the free energy (\ref{eq:frank}) must be supplemented by an additional term, 
\begin{equation}
F_\text{el}=-\frac{1}{2}\int\mathbf{D}\cdot\mathbf{E} \ \dd V.\label{eq:electricity}
\end{equation}
Here, $\vb{E}=-\nabla\phi$ is the electric field with $\phi$ being the electric potential, and $\mathbf{D}=\epsilon\mathbf{E}$. To find the equilibrium solution, we must minimize (\ref{eq:frank})+(\ref{eq:electricity}) as well as solve for the electric potential $\phi$. In Figure \ref{fig:tactoid}B, we show equilibrium solutions for increasing potential differences in two dimensions. The aspect ratio of the droplet is plotted in Fig. \ref{fig:tactoid}C showing elongation as the electric field is increased until, eventually, the solution becomes unstable. 

These two scenarios highlight the rich possibilities for shape change that arise in complex fluids. Many further extensions of the present examples are possible in \emph{Morpho}: Certain liquid crystal materials, for instance, adopt a spontaneously twisted structure and are known as cholesterics \cite{degennes_prost_1993,Carenza2022}. Such materials can easily be simulated by incorporating an additional term in (\ref{eq:frank}). Alternative theoretical formulations of liquid crystal elasticity \cite{degennes_prost_1993,Mottram2014} exist and can readily be used within the program. Additional physics can be included by formulation and inclusion of an appropriate energy functional. 

\subsection{Example 3: Swelling hydrogels}
\begin{figure*}
\centering
\includegraphics[width=\textwidth]{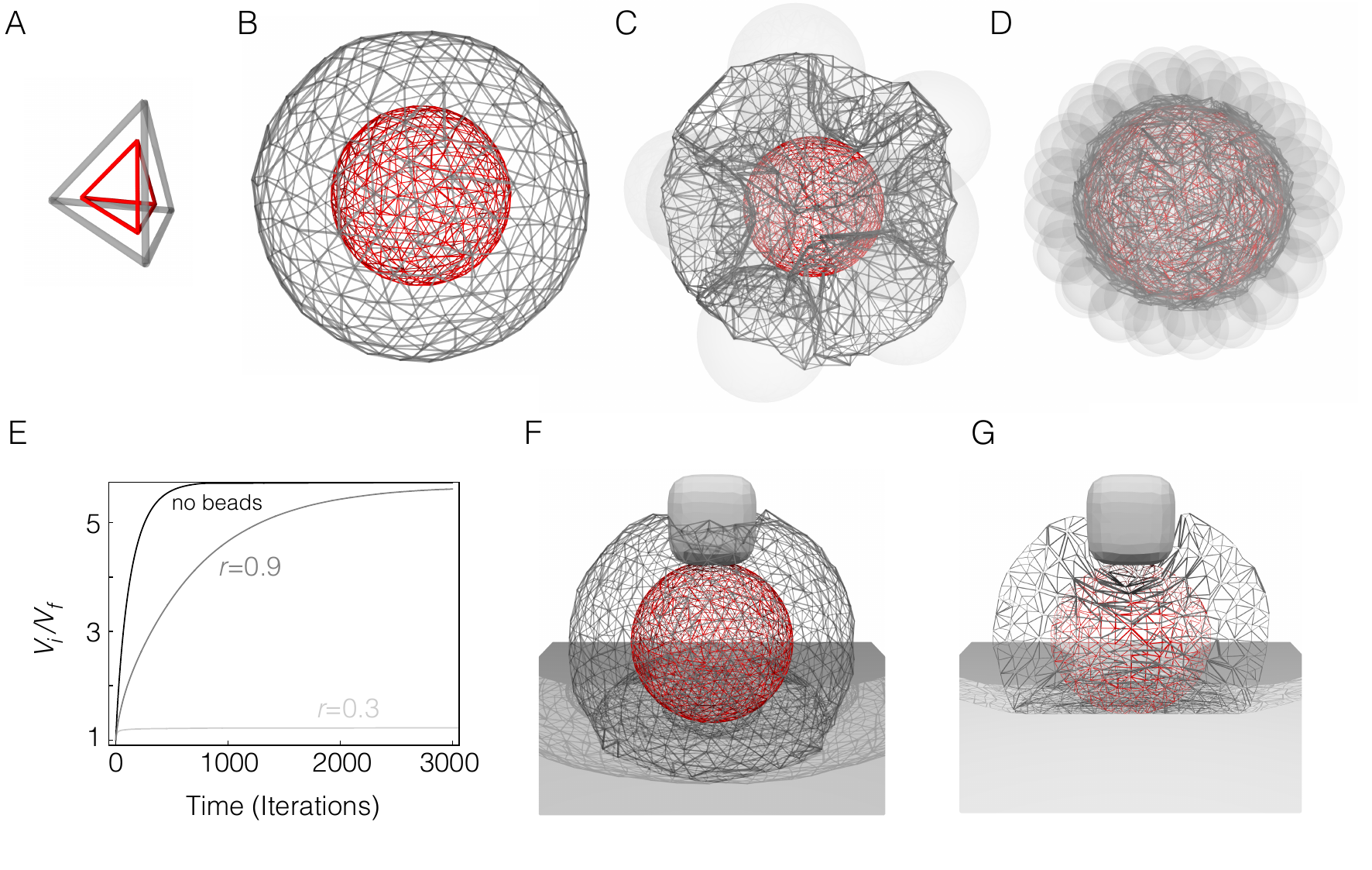}
\caption{\boldf{Modeling swelling hydrogels under confinement using tetrahedral meshes and level-set constraints.} In all sections, red colored mesh elements represent the initial state whereas the dark gray colored elements represent the final state. While the meshes are full 3D, only line elements at the boundary are displayed here for simplicity. Confinements / obstacles are shown in light gray. \textbf{A} A representative tetrahedral element, which forms the basis of the 3D Morpho meshes, before and after swelling. \textbf{B} A swollen spherical hydrogel without any constraints. \textbf{C}, \textbf{D} Hydrogel swelling in the presence of stationary hard sphere beads. \textbf{E} Relative change in volume as a function of time, under equal step-size gradient descent dynamics, for the unconstrained swelling in B and the constrained swelling in C and D. \textbf{F} Swelling in the proximity of a stationary impenetrable plane and a superellipsoid. \textbf{G} Planar slice of the mesh in F showing expansion around the superellipsoid. }
\label{fig:hydrogel}
\end{figure*}
Hydrogels comprise two components, a polymer network infiltrated by water. By adjusting the crosslinking density, fill fraction, etc. the rigidity of the gel can be adjusted across many orders of magnitude. Furthermore, these materials have an incredible capacity to absorb water while remaining intact \cite{Bertrand2016} making them amenable to a variety of different practical applications \cite{Ahmed2015}. Due to the inevitable shape change that occurs, continuum modeling of hydrogels is often restricted to simple geometries. Incorporating constraints presents further challenges \cite{Kang2010}. Here, we model a 3D hydrogel using a tetrahedral mesh. We adopt the Flory-Rehner formalism \cite{Quesada-Perez2011}, wherein the free energy $F$ is the sum of a mixing term and an entropic elastic term. We ignore the ionic contribution for now, but this can be easily incorporated. Equilibrium is defined by the balance of osmotic pressures, $\partial F / \partial \Ns=0$, with $\Ns$ being the number of solvent molecules. We equivalently write the free energy in terms of the volume fraction of the polymer $\phi$, which makes it easier to connect to other formalisms for the energy \cite{Rognes2009a}. We can compute this energy for any tetrahedron in the mesh, given its initial values for the volume $\Vi$ and volume fraction $\phii$ together with a reference volume fraction $\phi_0$ \cite{Quesada-Perez2011}. The resulting total energy is then minimized (see Methods) to obtain the equilibrium shape and size of the hydrogel.

Figure \ref{fig:hydrogel}A shows a representative tetrahedral element used in the computation. The finite-element approach allows us to realize arbitrary initial shapes of the hydrogel using a tetrahedral mesh. Here, we begin with a spherical mesh, where the result is a spherical shape of different size. The parameters of the Flory-Rehner functional determine the equilibrium volume fraction $\phieq$, and correspondingly, the expected change in volume $\Veq/\Vi$. Setting this ratio to be $\approx 6$, we perform gradient descent to reach the equilibrium. Figure \ref{fig:hydrogel}B shows the initial mesh in red and the final mesh in gray. 

\emph{Morpho} provides a convenient type of constraint, whereby vertices can be excluded from a region defined by the contours or level-sets of a scalar function. Using this, we can easily add confinement to the model computation. Inspired by recent beautiful experimental measurements \cite{Louf2021}, we introduce hard sphere beads surrounding the hydrogel. Figure \ref{fig:hydrogel}C shows the resulting swollen hydrogel. In Figure \ref{fig:hydrogel}D, we plot the ratio of the current volume to the initial volume for the case of unconfined and confined hydrogels from B and C, respectively. We can see a decrease in the swelling comparable to what is observed in \cite{Louf2021}. To demonstrate the flexibility in the constraint application, Figure \ref{fig:hydrogel}E and F show, respectively, the full 3D and a cross section of the swollen hydrogel in the presence of non-trivial confinements like a hard wall and a super-ellipsoidal bead together.

Lastly, such a simulation allows access to quantities that are harder to probe via experiments, such as the local strain during swelling, or the trend across varying material parameters. The programmable environment of \emph{Morpho} allows for such an analysis directly without the need for external tools.

\section{Discussion}

In this Article, we presented \emph{Morpho}, a programmable environment that aims to solve a broad class of shape optimization and shape-shifting problems. Using an explicit method as used in software such as the \emph{Surface Evolver}, we provide a programmable environment that goes beyond minimal surfaces and allows for domains \emph{and} fields to be minimized together in arbitrary number of dimensions. \emph{Morpho} provides additional features to facilitate mesh quality control: Leveraging the fact that shape optimization problems are often underdetermined, e.g. that the vertex positions for a minimal surface are uniquely specified only up to local tangential displacements, we employ auxiliary functionals that regularize the mesh elements and avoid clustering of vertices during optimization. 
 
We demonstrated \emph{Morpho}'s applicability to a variety of domains through three examples: In the first, minimizing the area of a closed membrane given a fixed volume, regularization permits the program to converge on the correct solution and improves the speed of convergence while the unregularized problem fails to converge. 

In a second example, we showed the combined minimization of the shape as well as associated fields in  liquid crystal tactoids. The customizable environment of \emph{Morpho} allows for automated adaptive refinement based on heuristics (or error estimators if available) such as energy density, enabling us to resolve liquid crystal defects known as boojums that occur in some tactoid solutions. We further demonstrated the solution of a multiphysics problem where we found shapes of tactoids elongated in the presence of an electric field, solving for the shape, liquid crystal director as well as the electric field.

Our final example showed constrained optimization by simulating hydrogels swelling under arbitrary confinement as well as \emph{Morpho}'s visualization capabilities, such as 3D ray-tracing provided by the \emph{povray} module and the 2D slice visualization made possible by the \emph{meshslice} module.

Beyond these examples, \emph{Morpho} could be used for many other domains of science and engineering, including wetting problems, liquid crystal membranes, thin shell problems, and so on. It could also be used for purely mathematical purposes such as computing arbitrary minimal surfaces like a 3-periodic gyroid and its variants. 

While the program as initially released is capable of solving a wide class of problems, we aim to further broaden the scope of \emph{Morpho} by developing a number of improvements. We plan to expand the range of discretizations available for both shape and fields, and incorporate improved optimization algorithms. Applications beyond optimization, e.g. non-equilibrium dynamics as is central to active biological materials \cite{Giomi2014,Maroudas-Sacks2020}, will be an important future target. 

\emph{Morpho} is entirely open-source under an MIT license and is provided with thorough documentation, readable from within the program in interactive mode or through a website as well as an extensive user manual. \emph{Morpho} is built with an automated testing suite to facilitate a high degree of reliability. While the program is already competitive with other software used in this space, we are presently working on GPU acceleration to further improve performance. We hope that the availability of our robust, well-tested open-source shape optimization software will benefit the soft matter physics community in particular and researchers interested in shape optimization and shape shiftin problems at large.

\section{Methods}

In the following subsections, we use \textbf{boldface} type to indicate
classes of object, \emph{italic} type to refer to modules available within \emph{Morpho} and \texttt{typewriter} font to refer to external programs. 

\subsection{Morpho overview}

\subsubsection{Model}

\emph{Morpho}'s organizing metaphor for shape optimization involves the following classes of object: 

\textbf{Meshes} may incorporate elements of various types. Meshes are \emph{graded}, i.e., can incorporate point-like, line-like, area-like elements, etc. that are stored separately. Each element is associated with an integer \emph{id} value. Connectivity information is stored in sparse matrices to facilitate use of efficient graph algorithms. To specify a mesh, it is only necessary to specify which vertices correspond to each element; additional connectivity information is automatically generated as needed. 

\textbf{Fields} are object collections that store floating point information, i.e., scalars or matrices, on elements of a mesh. A field may assign any number of quantities to each element, and may store a different number of quantities on elements of each grade. To facilitate fast operations, the underlying data store is monolithic and arithmetic operations implemented via BLAS etc. 

\textbf{Functional} objects correspond to terms in the energy functional and facilitate summation of contributions from individual elements. Given a mesh and any required fields, a Functional can return the total value corresponding to summing over elements in the mesh, a matrix of values for each element, the forces on each vertex and generalized forces on Field degrees of freedom. 

\textbf{Selections} are objects that represent selected portions of
a mesh. Each element can be selected or not. Selection objects can
then be used to achieve various effects, such as restricting a functional
to a particular subset of the mesh, or displaying a region of interest. 

\subsubsection{Scripting language}

\emph{Morpho} provides a simple but powerful object-oriented scripting language to describe and solve optimization problems. The syntax resembles other languages in the C family but has been kept small and clean to facilitate a low barrier to entry, inspired by the \texttt{Lox} language\cite{nystrom2021crafting}. Despite its simplicity, the language provides very good performance and supports many features expected in a modern dynamic language: modules, classes, closures and many common collection types such as lists, dictionaries and matrices both dense and sparse. Linear algebra utilizes efficient libraries such as \texttt{BLAS}\cite{blackford2002updated}, \texttt{LAPACK}\cite{lapack99} and \texttt{SuiteSparse}\cite{davis2006direct}.
The environment is highly extendable; a number of modules (all written in the \emph{Morpho} language) are included as standard and implement important \emph{Morpho} functionality. The user can easily add to these as they wish. The environment also provides good performance: scripts are compiled to an efficient bytecode for fast execution by a virtual machine, and the interpreter is among the fastest available.

\subsubsection{Optimization}

Optimization is provided by a separate \emph{optimize} module. The
problem to be solved---the target functional---is described by creating
an \textbf{OptimizationProblem} object followed by creating and adding
Functional objects. For example, the problem, 
\[
\text{Minimize}\ {\sigma\int_{\partial C}dA}+{\kappa\int_{\partial C}H^{2}dA}\ \text{s.t.}\ {\int_{C}dV}=V_{0},
\]
is represented in \textit{Morpho} through an \textbf{Area} object, a \textbf{MeanCurvatureSq}
object and a \textbf{VolumeEnclosed} object. The first two terms are
added to the \textbf{OptimizationProblem} using the \emph{addenergy}
method (with appropriate prefactors) and hence form part of the objective
function to be minimized. The volume constraint is added to the problem
using the \emph{addconstraint} method. Target values for constraints
may be specified, otherwise they are deduced from the initial state
of the system. 

Optimization is achieved by using \textbf{ShapeOptimizer} and \textbf{FieldOptimizer}
objects which act on a given \textbf{OptimizationProblem}. Optimizer
objects invoke the appropriate methods on each functional to calculate
the current total value of the target functional for the given configuration
as well as its gradient with respect to vertex positions or field
degrees of freedom. 

As a simple illustrative example, consider a single triangular element
with vertices $\mathbf{x}_{0}$, $\mathbf{x}_{1}$ and $\mathbf{x}_{2}$.
The area of the element is given by,
\[
A=\frac{1}{2}\left|(\mathbf{x}_{1}-\mathbf{x}_{0})\times(\mathbf{x}_{2}-\mathbf{x}_{1})\right|,
\]
from which we can readily compute $\nabla_{\mathbf{x}_{i}}A$, the
gradient with respect to each vertex position. Suppose all vertex
positions are stored sequentially in a single column vector $\mathbf{x}$;
then we may similarly define a gradient column vector $\mathbf{g}=\nabla A$
from the set of gradients with respect to individual positions, $\nabla_{\mathbf{x}_{i}}A,\forall i\in\{0,1,2\}$.
For a mesh consisting of many triangles, we can compute the total
area and the gradient of the area with respect to each vertex position
by summing up the contributions from each element. Each \textbf{Functional}
object provides methods \emph{total} and \emph{gradient} to return
exactly these objects.

\textbf{ShapeOptimizer} and \textbf{FieldOptimizer} provide a number
of standard algorithms for optimization\cite{boyd2004convex}, including \emph{gradient
descent}, where the vertex positions are updated, 
\[
\mathbf{x}_{i+1}=\mathbf{x}_{i}-\alpha\mathbf{g}_{i},
\]
for a fixed learning rate $\alpha$ and gradient of the target functional $\mathbf{g}_{i}$. Also available are \emph{backtracking
linesearches} and \emph{conjugate gradient} methods for unconstrained
optimization, with others forthcoming. Constraints are handled via
a projection method as follows. First, the gradient of the constraint
functional $\mathbf{h}_{i}$ is computed as for the target functional.
The descent direction is then computed by projecting out the component
of $\mathbf{g}_{i}$ in the direction of $\mathbf{h}_{i}$,
\[
\mathbf{p}_{i}=\mathbf{g}_{i}-\frac{\mathbf{g}_{i}\cdot\mathbf{h}_{i}}{\mathbf{h}_{i}\cdot\mathbf{h}_{i}}\mathbf{h}_{i}.
\]
Optimization proceeds using the descent direction $\mathbf{p}_{i}$
in place of the gradient. Following each iteration, re-projection steps
are then taken to re-satisfy the constraint. A tolerance may be set
to control the fidelity with which the constraint is maintained. A
number of possible convergence criteria may be set, including convergence
of the energy, the norm of the change in the position of the vertices,
etc. 

\subsubsection{Visualization and data interchange}

Simple but powerful support for visualization is provided by two modules. The \emph{plot} module enables the user to conveniently visualize
meshes, fields and selections. This uses the low-level module, \emph{graphics}, that represents a scene as a list of 3D graphics primitives, including spheres, cylinders, tubes, arrows, collections of simplices and text. The abstract representation empowers the user to easily create custom visualizations, and enables output to different formats. 

One such target is an included viewer application, \texttt{morphoview}, that is provided for convenient viewing. A further module, \emph{povray}, integrates with the widely used \texttt{povray} raytracer \cite{povray} to convert graphics objects to give easy access to raytraced output: All figures in this manuscript were generated by \emph{Morpho} and rendered with povray. 

More sophisticated visualizations can be produced using external applications like Paraview; to facilitate interchange with such programs \emph{Morpho} can export meshes and data in the commonly used VTK format.

\subsection{Application details}

\subsubsection{Minimal surfaces and membranes}

To solve the minimal surface problem, we construct an initial ellipsoidal
\textbf{Mesh} with aspect ratio $2$ by stretching an initially spherical
mesh. The problem is then set up as follows: An \textbf{OptimizationProblem}
object is defined with an \textbf{Area} object to compute the surface
area, and an \textbf{VolumeEnclosed} object added as a constraint.
Optimization is performed using a \textbf{ShapeOptimizer} object and
successive linesearches are performed until the relative change in
the energy is $<10^{-8}$. 

To incorporate regularization, a secondary \textbf{OptimizationProblem}
is created incorporating an \textbf{EquiElement} object and the \textbf{VolumeEnclosed}
object added as a constraint. A second \textbf{ShapeOptimizer} object
is used to perform the regularization. To perform optimization, we
now interleave linesearches on the surface tension and regularization
problems, finding that a ratio of around 2:1 leads to satisfactory
convergence. 

Visualization of the solutions is performed using the \emph{plot}
module. As shown in Fig.~\ref{fig:membranes}, without regularization the problem rapidly
gets stuck as elements near the cap shrink due to clustering of the
vertices. 

The membrane problem is solved using an initial disk \textbf{Mesh}
created with the \emph{meshgen} module. An \textbf{OptimizationProblem}
is created for the target problem including \textbf{Area} and \textbf{MeanCurvatureSq}
functionals, as well as a secondary \textbf{OptimizationProblem} with
an \textbf{EquiElement} object. During optimization, certain vertices
near the center of the disk are held at a fixed height using a level
set constraint imposed using a \textbf{ScalarPotential} object. Optimization
is performed using \textbf{ShapeOptimizer} objects with the target
and regularization regularization steps interleaved. Once convergence
is achieved, the tethered vertices are moved to a new height and the
shape re-optimized. Discrete refinement moves are also performed at
each of these new heights, where the mesh is equiangulated and large
triangles split into smaller triangles. 

\subsubsection{Liquid crystal tactoids}

To simulate the structure of the tactoid in \emph{Morpho}, we first construct
an initially spherical \textbf{Mesh }corresponding to the unit ball
$\left|\mathbf{x}\right|^{2}<1$ with \emph{Morpho}'s \emph{meshgen} module.
We also create a \textbf{Field} object to represent the nematic director $\mathbf{n}$ with an initially uniform configuration
and a \textbf{Selection} object corresponding to the boundary of the
mesh. 

An \textbf{OptimizationProblem} object is then defined in Morpho by
creating and adding the following functional objects: a \textbf{Nematic}
object provides nematic elasticity; an \textbf{Area} object is used
to evaluate the surface area on the boundary; an \textbf{AreaIntegral}
is used to evaluate the anchoring energy. For two dimensional tactoids,
the anchoring energy must be replaced with an equivalent \textbf{LineIntegral}.
The director length constraint is imposed using a \textbf{NormSq}
object that calculates the norm-squared of every entry in the field;
this is added to the problem as a local constraint on the field \textbf{$\mathbf{n}$}.
Finally, a \textbf{Volume} object is used to evaluate the volume of
the mesh and is added to the optimization problem as a global constraint. 

Having set up the problem, separate \textbf{FieldOptimizer} and \textbf{ShapeOptimizer}
objects are created to optimize the field and shape individually. We
use both these objects to perform conjugate gradient steps on $\mathbf{n}$
and the mesh respectively. We empirically find that interleaving field
and shape optimization steps, with $\approx4\times$ more field optimization
steps, leads to reasonable convergence for the parameters considered.
An energy convergence criterion is used whereby the optimization problem
is considered to have converged when the relative change in the energy
is $<10^{-8}$. Having obtained a coarse solution, we perform adaptive
refinement by splitting elements with elastic energy $>1.5\times$
the mean using a \textbf{MeshRefiner} object from the \emph{meshtools} module.
We then optimize the refined solution as before. 

Visualization of the solutions obtained is performed with the \emph{plot}
module. We write a custom function to represent the nematic configuration
as a cylinder at every vertex and combine this with the surface of
the tactoid visualized with the \emph{plot} module; the combined output
is then rendered using the \emph{povray} module. 

\subsubsection{Swelling hydrogels}

We use a thermodynamic theory of swelling hydrogels \cite{fernandez-nieves_2011,Quesada-Perez2011,Flory1943,Flory1943a}. Here, a binary polymer-solvent mixture is considered, with $\Np$ and $\Ns$ being the number of polymer and solvent molecules respectively and $\nup$ and $\nus$ being the corresponding molar volumes. Equilibrium is reached when the chemical potential of the solvent is balanced inside and outside the hydrogel. Equivalently, this can be seen as minimizing the change in Helmholtz free energy $\Delta F$ with respect to the number of solvent molecules $\Ns$. This change in the free energy, under a separability approximation, can we written as 
\begin{equation}
    \Delta F = \Delta F_{\text{mix}} + \Delta F_{\text{el}}. 
\end{equation}
The first term is the Flory-Huggins mixing contribution \footnote{Note that this expression is valid in the assumption that the number of polymer repeat units is much larger than 1.}:
\begin{equation}
\label{eq:flory-huggins}
     \Delta F_{\text{mix}} = \Ns \kb T \left[ \ln (1-\phi) + \chi \phi \right].
\end{equation}
Here, $\phi$ is the volume fraction of the hydrogel, $\chi$ the Flory-Huggins mixing parameter, $\kb$ the Boltzmann constant and $T$ the temperature. Now, the osmotic pressure contribution from this energy is 
\begin{equation}
    \Pi_{\text{mix}} = -\frac{\NA}{\nus} \pdv{\Delta F_{\text{mix}}}{\Ns},
\end{equation}
with $\NA$ being Avogadro's number. Note that $\phi$ is dependent on $\Ns$:
\begin{equation}
    \phi = \frac{\Np \nup}{(\Np\nup + \Ns\nus)}.
    \label{eq:phi}
\end{equation}
Using this relation, we can recover the osmotic pressure
\begin{equation}
    \Pi_{\text{mix}} = -\frac{\NA}{\nus} \pdv{\Delta F_{\text{mix}}}{\Ns} = -\frac{\NA \kb T}{\nus} \left[ \phi + \ln (1-\phi) + \chi \phi^2 \right].
\end{equation}

Note that in the literature, this osmotic pressure is sometimes expressed in terms of an `effective diameter' $\alpha$ of the solvent molecule \cite{Louf2021}:
\begin{equation}
    \Pi_{\text{mix}} = -\frac{\kb T}{\alpha^3} \left[ \phi + \ln (1-\phi) + \chi \phi^2 \right],
\end{equation}
which implies $\alpha^3 = \nus / \NA$.

The elastic contribution to the free energy, $\Delta F_\text{el}$, is given by the Flory-Rehner elastic energy \cite{Quesada-Perez2011,Flory1943,Flory1943a},
\begin{equation}
    \Delta F_\text{el} = -\frac{3 \Nc \kb T}{2} (\ln \beta - \beta^2 + 1),
\end{equation}
where $\Nc$ is the number of chains in the network and $\beta = (V/V_0)^{1/3} = (\phi_0/\phi)^{1/3}$, with $V_0$ and $\phi_0$ being the volume and volume fraction respectively in the `reference' state \cite{Quesada-Perez2011}. 

In this work, we consider a 3D polymer hydrogel in a solvent bath at a fixed temperature $T$, where the volume fraction of the polymer can vary over space. Hence, we want to think about a free energy \emph{density} $\Delta f_{\text{mix}} (\vb{x})$, in terms of a spatially varying field $\phi(\vb{x})$. If this space is discretized using tetrahedra, it is useful to consider the expression \eqref{eq:flory-huggins} for a single tetrahedral element. The energy density locally at a point $\vb{x}$ \emph{in the deformed frame of reference} will be Eq.~\eqref{eq:flory-huggins} evaluated at $\vb{x}$ divided by the volume of the element. Since this volume would also be given by $\Np \nup + \Ns \nus$, we have

\begin{align}
    \Delta f_{\text{mix}} &= \frac{\Ns}{\Np \nup + \Ns \nus} \kb T \left[ \ln (1-\phi) + \chi \phi \right] \\
    &= \frac{(1-\phi)}{\nus} \kb T \left[ \ln (1-\phi) + \chi \phi \right].
\end{align}
In terms of $\alpha$, this would be
\begin{equation}
    \Delta f_{\text{mix}} = \frac{\kb T}{\NA \alpha^3}  \left[ (1-\phi)\ln (1-\phi) + \chi \phi (1-\phi) \right].
\end{equation}

Similarly for the elastic energy, we can compute the free energy by dividing by the volume. In this work, we assume that the chains are uniformly distributed throughout the hydrogel, so $\Nc$ doesn't depend on $\vb{x}$, but the functionality in \emph{Morpho} can be easily extended to allow a spatially varying initial $\Nc$. 

Finally, we note that it follows from Eq.\eqref{eq:phi} that minimizing w.r.t. $\Ns$ is equivalent to minimizing w.r.t. $\phi$. We thus connect to an equivalent formalism \cite{Rognes2009a} and implement the minimization w.r.t. $\phi$ in \emph{Morpho}. It can be seen that we have three non-dimensional parameters, namely, the Flory-Huggins mixing parameter $\chi$, the relative strength of the elastic energy to the mixing energy $\Nc \alpha^3/V_0$ and the reference volume fraction $\phi_0$ \cite{Quesada-Perez2011}. Given an initial value of $\phi$, we can vary these parameters to change the minima of the overall free energy, thus tuning the swelling ratio  (since $\phi_\text{final}$ affects $V_\text{final}$).   

To compute the structure of the hydrogel in \emph{Morpho}, we again start by constructing an initially spherical \textbf{Mesh} corresponding to the unit ball $\left|\mathbf{x}\right|^{2}<1$ with morpho's \emph{meshgen} module. An \textbf{OptimizationProblem} object is then defined and a \textbf{Hydrogel} functional, implementing the above discussed free energy density, is added to it. For hard confinements, we define level-set constraints corresponding to the objects (spheres, ellipsoids, planes, etc.) through the \textbf{ScalarPotential} object from the \emph{functionals} module. A \textbf{ShapeOptimizer} object is then created to optimize the shape. We perform gradient descent with a fixed step size to simulate inviscid dynamics of the swelling. A \textbf{Volume} object is used to keep track of the volume of the hydrogel during relaxation. 

To initialize the positions of the hard spheres for Figure \ref{fig:hydrogel}C, we define a dummy shell mesh with radius $R + R_\text{bead}$ with $N_\text{bead}$ number of vertices placed randomly. We first confine the vertices to lie on the shell by using a \textbf{ScalarPotential} object. We then define an electrostatic repulsive pairwise interaction between the vertices using a \textbf{PairwisePotential} object from the \emph{functionals} module, thus proceeding to solve the Thomson problem. The resulting mesh vertex positions are used as the sphere centers for the level set constraints. We thus get equidistantly packed spheres on the outer shell. 

All 3D visualizations in Figure \ref{fig:hydrogel} are made using the \emph{povray} module. The superellipsoid constraint in \ref{fig:hydrogel}E and \ref{fig:hydrogel}F is shown by constructing an equivalent mesh using the \emph{meshgen} module and plotting its facets. Similarly, the plane is plotted by defining an equivalent planar mesh, while the slice in \ref{fig:hydrogel}F is plotted using the \emph{meshslice} module.

\section*{Data availability}

Source data are provided with this paper.

\section*{Code availability}

The \emph{Morpho} application can be found at \url{https://github.com/Morpho-lang/morpho}.
A manual is included in the repository, together with a number of
examples. Source code for all examples shown in this publication can be found at \url{https://github.com/Morpho-lang/morpho-paper}. All code is released under an
open source MIT License. 

\section*{Acknowledgements}

This material is based upon work supported by the National Science Foundation under Grant No. ACI-2003820. TJA thanks the many people who have used various versions of the program or otherwise contributed to the project --- Abigail Wilson, Allison Culbert, Andrew DeBenedictis, Badel Mbanga, Chris Burke, Ian Hunter, Mathew Giso, Matthew S. E. Peterson, and Zhaoyu Xie.

\end{document}